\documentclass[aps,superscriptaddress,preprintnumbers,amsmath,amssymb,reprint,prl]{revtex4-2}
\usepackage{graphicx}
\usepackage{color}
\usepackage{xcolor}
\usepackage{hyperref}
\hypersetup{
    colorlinks=true,
    linkcolor=blue,
    filecolor=black,      
    urlcolor=blue,
    citecolor=blue,
}
\usepackage{verbatim}

\usepackage{soul}

\usepackage{lineno}

\newcommand{\beginSI}{%
        \setcounter{table}{0}
        \renewcommand{\thetable}{M\arabic{table}}%
        \setcounter{figure}{0}
        \renewcommand{\figurename}{Extended Data Fig.}
        \renewcommand{\thefigure}{\arabic{figure}}
        \setcounter{equation}{0}
        \renewcommand{\theequation}{M\arabic{equation}}%
        \renewcommand{\thesection}{S\arabic{section}}%
     }

\begin{document}


\title{Programming strain-stiffening in soft composites via structural memory near jamming }

\author{Yiqiu Zhao}
\thanks{These two authors contributed equally}
\affiliation{Department of Physics, The Hong Kong University of Science and Technology, Hong Kong SAR, China}

\author{Deng Pan}
\thanks{These two authors contributed equally}
\affiliation{Institute of Theoretical Physics, Chinese Academy of Sciences, Beijing 100190, China}

\author{Yiming Pang}
\affiliation{Department of Physics, The Hong Kong University of Science and Technology, Hong Kong SAR, China}

\author{Jonathan Bar\'{e}s}
\affiliation{LMGC, Universit\'{e} de Montpellier, CNRS, Montpellier, France}

\author{Chang Xu}
\affiliation{Department of Physics, The Hong Kong University of Science and Technology, Hong Kong SAR, China}

\author{Che Liu}
\affiliation{Department of Physics, The Hong Kong University of Science and Technology, Hong Kong SAR, China}

\author{Haitao Hu}
\affiliation{Department of Physics, The Hong Kong University of Science and Technology, Hong Kong SAR, China}

\author{Yuliang Jin}
\email{yuliangjin@mail.itp.ac.cn}
\affiliation{Institute of Theoretical Physics, Chinese Academy of Sciences, Beijing 100190, China}

\author{Qin Xu}
\email{qinxu@ust.hk}
\affiliation{Department of Physics, The Hong Kong University of Science and Technology, Hong Kong SAR, China}

\date{\today}

\begin{abstract}   
    Soft composite solids, comprising discrete inclusions embedded within a compliant matrix, are emerging candidates for engineering synthetic tissues and soft robotic materials. Current strategies for controlling their nonlinear mechanics, such as strain-stiffening, have primarily relied on the nonlinear elasticity of polymer matrices. Although direct contacts between inclusions may enhance stiffening responses at high densities, the role of the non-equilibrium and history-dependent nature of disordered contact networks in composite mechanics remains unexplored. In this work, by applying a mechanical training protocol near a shear-jamming phase boundary, we demonstrate that the  structural memory encoded in contact networks drives a crossover from granular-like to biopolymer-like strain stiffening. Simulations of a coarse-grained composite model reveal that this biopolymer-like mechanical response emerges from enhanced non-affine reconfigurations of nearly-jammed contact networks. Without relying on matrix nonlinearity, we establish a design strategy that leverages non-equilibrium memory effects intrinsic to granular systems to achieve highly programmable strain-stiffening in soft composites.
\end{abstract}

\maketitle
\noindent

Soft solids, ranging from polymeric gels to biological tissues, often exhibit nonlinear mechanics that are critical for their tailored functionalities. In living systems, the strain-stiffening response of soft tissues helps preserve the structural integrity of organs~\cite{Licup2015_pnas,Sharma2016_NatPhys,Burla2019_NatPhys,huang2022_prl}.  In synthetic contexts, multi-component soft composites comprising micro-inclusions embedded in a compliant matrix have emerged as promising  candidates to replicate such nonlinear stiffening responses~\cite{fang2020_matter, Xie2021_NatCom, subramaniam2024_giant,Xue2025_nc}. 
With the growing use of soft composites in tissue engineering~\cite{Guimaraes2020_nrm,song2021_jap,Song2025_np}, soft robotics~\cite{Hu2018_nature,Deng2020_nc,Bao2025_nature}, and wearable devices~\cite{Koydemir2018_Review,Tanriverdi2025_nc,Pu2025_nature}, establishing design principles for programming their nonlinear mechanics has become increasingly desirable.

For tissue-mimetic composites consisting of biopolymer networks, embedded inclusions tend to locally amplify strain and stress within the polymeric matrix, resulting in enhanced macroscopic strain-stiffening~\cite{van_Oosten2019_Nature}. While this mechanism, relying on the matrix's nonlinear elasticity, effectively captures the mechanics of loosely dispersed composites, it may break down in the dense limit. Recent studies have revealed that the percolated contact networks of inclusions can govern the composite mechanics~\cite{Shivers2020_PNAS,zhao2024_nc}, providing a different design principle based on jamming criticality of dense inclusions~\cite{xu2026_nm}.  

Systems near jamming are known to retain memories of their preparation history~\cite{keim2019_rmp,Zhao2022_PRX,candela2023_prl,ong2024_prx}. In discrete granular materials, simulations have shown that pre-engineered contact networks can shift shear-jamming boundary~\cite{Kumar2016_gm,Jin2021_pnas,kawasaki2024_prl} and thus modulate the elasticity of jammed states~\cite{pan2023_pnas_n}. These non-equilibrium features offer a potential route for tuning the mechanics of disordered contact networks. However, it remains unclear how this inherent granular memory influences the nonlinear mechanics of soft composites.

In this work, we employ a training protocol that prepares the inclusion configurations according to a history-dependent shear-jamming plane. By harnessing the competition between non-affine elastic energy and contact networks, we uncover a crossover from granular-like to biopolymer-like stiffening, characterized by a transition across distinct stiffening exponents. These findings suggest a design principle for  strain-stiffening in soft composites by controlling the structural memory of inclusions, rather than the matrix’s nonlinear elasticity.

\begin{figure*}[t]
    \centering
    \includegraphics[width = \textwidth]{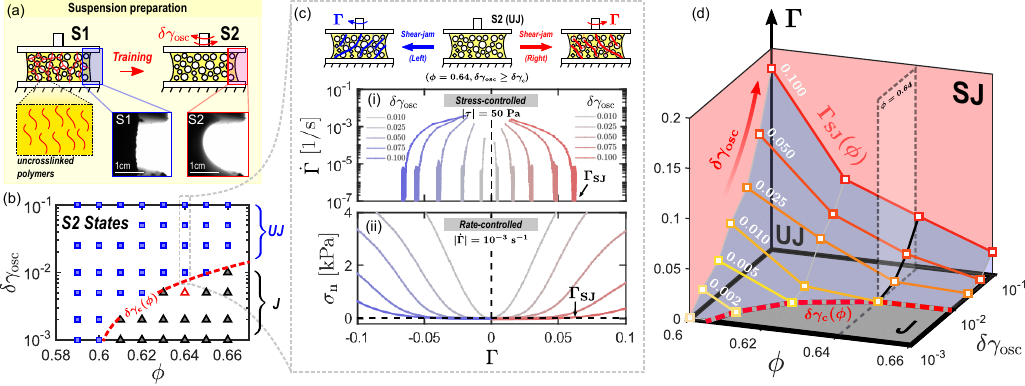}
    \caption{
        {\bf{Mechanical training of particle configurations.}}
        (a) Schematic of the training protocol. An as-prepared PS-PDMS suspension (S1) is trained under an oscillatory shear (amplitude $\delta\gamma_{\rm osc}$). The resulting suspension (S2) is unjammed but maintains a trained structural memory. Inset: Snapshots of a PS-PDMS suspension ($\phi = 0.62$) before and after training with $\delta\gamma_{\rm osc} = 0.10$. 
        (b) Phase diagram of trained suspensions (S2) for different $\phi$ and $\delta\gamma_{\rm osc}$. The red dashed line indicates the critical strain amplitude $\delta\gamma_{\rm c}(\phi)$, marking the boundary between jammed (J) and unjammed (UJ) states. The red open triangle indicates the training strain $\delta\gamma_{\rm osc} = 0.005$ near $\delta\gamma_c$ for $\phi = 0.64$. 
        (c) Re-jamming of trained suspensions (S2). For a PS-PDMS suspension ($\phi = 0.64$) trained with different $\delta\gamma_{\rm osc}$,  shear rate ($\dot\Gamma$) is plotted against shear strain ($\Gamma$) under a constant shear stress $\tau = 50$~Pa (sub-panel (i)), and normal stress ($\sigma_n$) is plotted against $\Gamma$ under a constant shear rate $\dot\Gamma=10^{-3}$~s$^{-1}$ (sub-panel (ii)). The shear jamming strain $\Gamma_{\rm SJ}$ is identified by the sharp drop in shear rate ($\dot\Gamma$) to zero.  
        (d) Plot of $\Gamma_{\rm SJ}$ against $\phi$ and $\delta\gamma_{\rm osc}$, demonstrating a history-dependent shear jamming transition (SJ).
        For each $\phi$, $\Gamma_{\rm SJ} > 0$ only exists when $\delta\gamma_{\rm osc}  > \delta \gamma_{\rm c}(\phi)$. At $\delta\gamma_{\rm osc} \approx \delta \gamma_{\rm c}(\phi)$, the suspensions are marginally jammed with $\Gamma_{\rm SJ} = 0$.   
           }
    \label{fig:training}
\end{figure*}

\subsection{\bf Mechanical training of  particle configurations
}

To endow soft composites with structural memory of inclusions, we began by training the disordered packing
of inclusions in a polymeric melt. After polymer crosslinking, these pre-engineered particle configurations were preserved in the final composite solids.  As illustrated in Fig.~\ref{fig:training}(a), suspensions of 29~$\mu$m polystyrene (PS) microspheres in polydimethylsiloxane (PDMS) melts were initially prepared at a high volume volume fraction ($\phi > 0.58$). Due to the large strain induced by mixing, the as-prepared suspensions (state S1) were always jammed. Through an oscillatory shear with a constant frequency $\omega = 100$ rad/s and amplitude $\delta \gamma_{\rm osc}$ applied for $180$~s, the PS–PDMS suspensions were mechanically relaxed and trained into a distinct state (state S2). Figure~\ref{fig:training}(b) shows that whether the trained suspensions (S2) are jammed or unjammed depends on $\phi$ and $\delta \gamma_{\rm osc}$, with the two different S2 states separated by a phase boundary $\delta\gamma_{\rm c}(\phi)$ (red dashed line). 
A similar phase diagram was numerically predicted in previous simulations of cyclically sheared repulsive spheres~\cite{das2020_pnas}.

Starting from a given $S1$ state, a gradual increase in $\delta\gamma_{\rm osc}$ beyond $\delta\gamma_{\rm c}$ triggers an unjamming transition, which is characterized by the relaxation of both global and local stresses in suspensions (Extended Data Fig.~\ref{fig:extend:unjamming} and Supplementary Video 1). The unjammed states prepared with $\delta\gamma_{\rm osc} > \delta\gamma_c (\phi)$ are not structurally identical. Instead, they carry the memories encoded by $\delta\gamma_{\rm osc}$, which can be subsequently read out through a shear re-jamming process. For example, Fig.~\ref{fig:training}(c) shows the responses of suspensions ($\phi=0.64$) trained with various $\delta\gamma_{\rm osc}$ to a unidrectional shear ($\Gamma$). Under a constant shear stress $\tau = 50$~Pa, the shear rate $\dot\Gamma$ drops sharply to zero at a critical strain $\Gamma_{\rm SJ}$ in both shear directions (subpanel (i)). Under a constant shear-rate $\dot\Gamma =10^{-3}$~s$^{-1}$, $\Gamma_{\rm SJ}$ marks the onset at which the normal stress $\sigma_{\rm n}$ turns positive from zero (subpanel (ii)). Thus, $\Gamma_{\rm SJ}$ characterizes the shear strain that required to re-jam the trained suspensions. Since $\Gamma_{\rm SJ}$ increases monotonically with $\delta\gamma_{\rm osc}$ and vanishes as $\delta\gamma_{\rm osc}$ approaches $\delta\gamma_c$,  it quantifies structural memory when $\delta\gamma_{\rm osc} \geq 0.01$.  In contrast, at the phase boundary $\delta\gamma_{\rm osc} = 0.005\approx \delta\gamma_c (\phi = 0.64)$ (the red open triangle in Fig.~\ref{fig:training}(b)), the as-prepared suspensions were marginally jammed without shear, implying $\Gamma_{\rm SJ} =0$.


Figure~\ref{fig:training}(d) presents $\Gamma_{\rm SJ}$ as a function of $\phi$ and $\delta\gamma_{\rm osc}$. For each $\phi$ between 0.60 and 0.66, starting from the phase boundary $\delta\gamma_{\rm osc} \approx \delta\gamma_c (\phi)$, $\Gamma_{\rm SJ}$ increases monotonically from zero with $\delta\gamma_{\rm osc}$.  The contour plot $\Gamma_{\rm SJ}(\phi, \delta\gamma_{\rm osc})$ reveals a history-dependent, non-equilibrium jamming plane for the PS-PDMS suspensions, a feature previously explored only in simulations~\cite{Kumar2016_gm,Jin2021_pnas,kawasaki2024_prl}.
In a vertical plane at a fixed volume fraction ($\phi$), $\Gamma_{\rm SJ}(\delta \gamma_{\rm osc})$ indicates a training-dependent shear jamming transition.
Without loss of generality, our subsequent experiments focused exclusively on the precursor suspensions with a fixed $\phi = 0.64$, trained in the regime $\delta\gamma_{\rm osc} \geq \delta\gamma_c\approx0.005$ (the region bounded by the dashed line in Fig.~\ref{fig:training}(d)).

\begin{figure*}[!t]
    \centering
    \includegraphics[width=\textwidth]{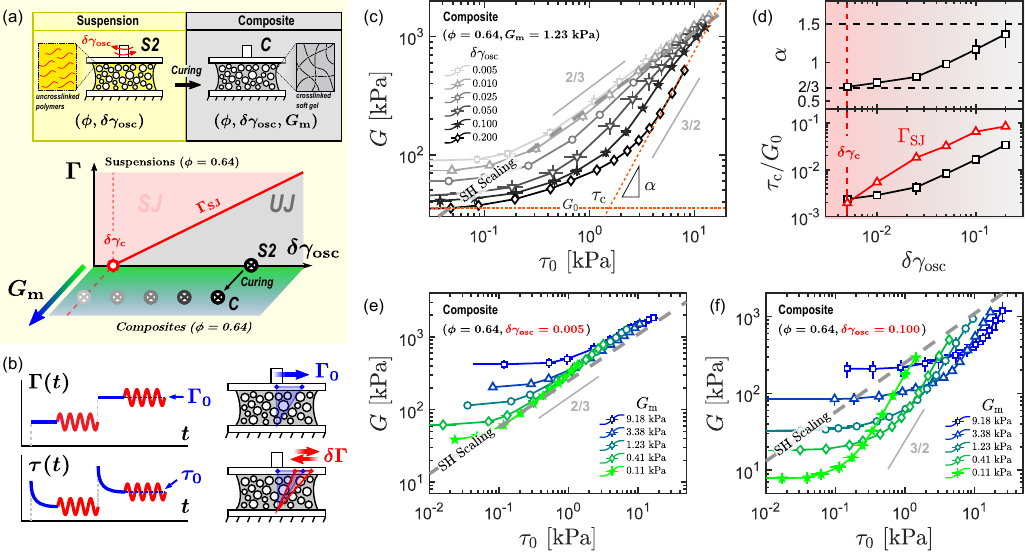}
    \caption{{\bf Memory-controlled strain stiffening.}
    (a) Schematic of the curing process of soft composites. Crosslinking the PDMS melts converts the trained suspensions (state S2) into soft composites with engineered structural memories (state C). For a fixed volume fraction $\phi=0.64$, the mechanics of soft composites is controlled by not only the measurement strain ($\Gamma$) and training oscillatory strain ($\delta\gamma_{\rm osc}$), but also the shear modulus of the matrix ($G_{\rm m}$). 
    (b) A step strain $\Gamma_0$ is applied to soft composites, followed by a superimposed oscillatory shear, with a strain amplitude $\delta\Gamma = 0.1~\%$ and oscillation frequency $\omega = 1.27$~rad/s. The resulting shear stress oscillates around a mean value $\tau_0$. The quasi-static shear modulus ($G$) is quantified using the low-frequency storage modulus. 
    (c) Plots of $G$ versus $\tau_0$ for PS-PDMS composites ($\phi=0.64$ and $G_{\rm m} = 1.23$~kPa) trained with various $\delta\gamma_{\rm osc}$. 
    (d) Plots of the stiffening exponent $\alpha$ (top panel) and the ratio $ \tau_{\rm c}/G_0$ (bottom panel)  against $\delta\gamma_{\rm osc}$ for the stiffening traces shown in (c). 
    (e) Plot of $G$ versus $\tau_0$ for PS–PDMS composites ($\phi = 0.64$) with varying $G_{\rm m}$, which are all trained by $\delta\gamma_{\rm osc} = 0.005\approx \delta\gamma_{\rm c}$.
    (f) Plot of $G$ versus $\tau_0$ for stiffening curves for PS–PDMS composites ($\phi = 0.64$) with varying $G_{\rm m}$, which are all trained by $\delta\gamma_{\rm osc} = 0.10> \delta\gamma_{\rm c}$. The grey dashed lines in (c), (e), and (f) indicate the shear hardening (SH) scaling predicted by Eq.~\ref{eq:sh}. The error bars in (c)–(f) represent the standard errors of the mean from at least three independent measurements.}
    \label{fig:memory}
\end{figure*}

\subsection{Memory-controlled strain stiffening}

The trained suspensions were encoded with a structural memory of particle networks in the absence of an elastic matrix. To preserve these engineered configurations in the resulting soft composite solids, we then crosslinked PDMS solvent into a gel {\em in-situ}, while fixing the shear plates (Fig.~\ref{fig:memory}(a)). The crosslinking density was systematically varied to tune the shear modulus of the PDMS matrix ($G_{\rm m}$) over two orders of magnitude, ranging from 0.1~kPa to 10~kPa The composite mechanics depended on both the training protocol of contact networks ($\delta\gamma_{\rm osc}$) and the matrix stiffness ($G_{\rm m}$). 

The nonlinear mechanics of soft composites were characterized by superposition rheology~\cite{gardel2004_prl,lin2010_prl}. As illustrated in Fig.~\ref{fig:memory}(b), the applied shear strain ($\Gamma_0$) on a soft composite is increased stepwise. Upon reaching the equilibrium shear stress ($\tau_0$) at each step, a small shear oscillation (strain amplitude $\delta\Gamma  = 0.1~\%$) at a low frequency $\omega = 1.27$~rad/s is superimposed. The storage modulus provides the shear modulus ($G$) of the composites under varying shear stresses $\tau_0$ (Fig.~S1).

Figure~\ref{fig:memory}(c) shows $G(\tau_0)$ for soft composites prepared with constant material parameters ($\phi = 0.64$ and $G_{\rm m}= 1.23$~kPa) but subjected to varying training amplitudes $\delta\gamma_{\rm osc}$. As $\delta\gamma_{\rm osc}$ increases from 0.005 to 0.20, the strain-stiffening regimes, characterized by a power-law scaling $G\sim \tau_0^\alpha$, becomes more pronounced, with the stiffening exponent $\alpha$ transitioning from $2/3$ to $3/2$ (top panel, Fig.~\ref{fig:memory}(d)). Over the same range of $\delta\gamma_{\rm osc}$, the onset strain, defined as the ratio of the onset stress ($\tau_c$) to the shear modulus in the low-stress plateau ($G_0$), rises by one order of magnitude (bottom panel in Fig.~\ref{fig:memory}(d)). Notably, the variation of $\tau_c/G_0$ with $\delta\gamma_{\rm osc}$ qualitatively aligns with the history-dependent re-jamming shear strain $\Gamma_{\rm SJ}(\delta \gamma_{\rm osc})$, suggesting a central role for shear-jamming transitions in governing the stiffening response. 

For $\delta\gamma_{\rm osc} = 0.005 \approx \delta\gamma_{\rm c}$, the embedded PS particles are marginally jammed. As $G_{\rm m}$ is systematically increased from 0.11 to 9.18~kPa, all traces of $G(\tau_0)$ within the stiffening regime collapse onto the same scaling, $G\sim \tau_0^{2/3}$, independent of $G_{\rm m}$ (Fig.~\ref{fig:memory}(e)). This master curve is consistent with the shear hardening observed in simulations for jammed granular solids (gray dashed line):
\begin{equation}\label{eq:sh}
    G=a_{\rm sh} (E^*_{\rm p})^{1-\alpha_{\rm sh}}\tau_0^{\alpha_{\rm sh}}
\end{equation}
where $a_{\rm sh} = 1.28$ is a dimensionless fitting parameter,
$E_{\rm p}^*=E_{\rm p}/2(1-\nu_{\rm p}^2) = 2.3$ GPa represents the plane-strain elastic modulus of the PS particles (with $E_{\rm p}$ and $\nu_{\rm p}$ the Young's modulus and the Poisson's ratio of the PS particles), and $\alpha_{\rm sh} = 0.64$ corresponds to the shear hardening exponent for jammed Hertzian spheres (Fig.~S2). In stark contrast, for $\delta\gamma_{\rm osc} = 0.1\gg\delta\gamma_{\rm c}$, the composites exhibit pronounced strain-stiffening with an exponent $\alpha \approx 3/2$ (Fig.~\ref{fig:memory}(f)), a signature characteristic of biopolymer networks~\cite{gardel2004_science,Broedersz2014_RMP,meng2016_soft,Prince2024_bm}. Since the values of $G(\tau_0)$ in the stiffening regimes depend significantly on the matrix modulus ($G_{\rm m}$), the enhanced stiffening exponent ($\alpha \approx 3/2$) stems from a physical mechanism fundamentally distinct from the granular shear-hardening ($\alpha \approx 2/3$) at $\delta\gamma_{\rm osc} \approx \delta\gamma_{\rm c}$. 

\begin{figure}[t]
    \centering
    \includegraphics[width= 87 mm]{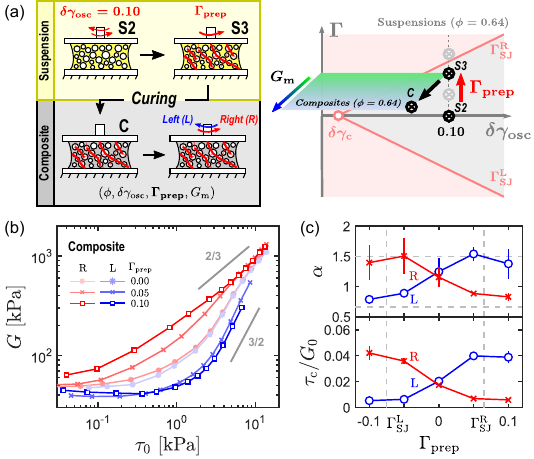}
    \caption{{\bf Asymmetric stiffening responses.}
        (a) Schematic illustration of the fabrication of soft composites with asymmetric stiffening responses. In addition to the oscillatory shear ($\delta\gamma_{\rm osc}$), a unidirectional pre-strain ($\Gamma_{\rm prep}$) is further applied to the suspensions. In the $\delta\gamma_{\rm osc}-\Gamma_{\rm prep}$ phase diagram, the resulting contact networks of inclusions lie closer to the shear-jamming boundary along the direction of $\Gamma_{\rm prep}$ than along the opposite direction. 
        (b)  Plots of $G$ against $\tau_0$  in both ``Right" (R) and ``Left" (L) directions, measured for three soft composites prepared with different pre-strains $\Gamma_{\rm prep} = 0, 0.05, 0.10$. 
        (c)  Plots of stiffening exponent $\alpha$ and the onset strain $\tau_{\rm c}/G_0$ against $\Gamma_{\rm prep}$, where $\Gamma_{\rm SJ}^{\rm R}$ and $\Gamma_{\rm SJ}^{\rm L}$ denote the shear jamming boundaries in R- and L- directions, respectively. The error bars represent the fitting uncertainties.     
    }
    \label{fig:asymmetric}
\end{figure}

We further demonstrate asymmetric stiffening responses within a single composite by exploiting the directional memory of shear-jammed contact networks. As illustrated in Fig.~\ref{fig:asymmetric}(a), a PS-PDMS suspension ($\phi=0.64$) is first subjected to an oscillatory shear with $\delta\gamma_{\rm osc} = 0.10$ for training (reference state S2). A unidirectional shear strain $\Gamma_{\rm prep}$ is subsequently applied to induce contact asymmetry (state S3). Upon curing the matrix at state S3, the resulting composite retains the memory induced by both $\delta\gamma_{\rm osc}$ and $\Gamma_{\rm prep}$. In the $\Gamma-\delta\gamma_{\rm osc}$ phase diagram, the embedded contact network is thus positioned closer to the shear-jamming boundary along the direction of $\Gamma_{\rm prep}$ than in the opposite direction.

To probe this structural asymmetry experimentally, we prepared three composite samples using $\Gamma_{\rm prep}=0, 0.05$, and $0.10$, while maintaining $\phi = 0.64$, $G_{\rm m}= 1.23$~kPa, and $\delta\gamma_{\rm osc} = 0.10$. Figure~\ref{fig:asymmetric}(b) presents the directional strain-stiffening responses, where ``Right'' (R) refers to the same direction as $\Gamma_{\rm prep}$ and ``Left'' (L) refers to the opposite direction. While the plot of $G(\tau_0)$ remains symmetric with respect to shear for $\Gamma_{\rm prep} = 0$, the difference between the R and L measurements becomes significant when $\Gamma_{\rm prep} >0$. For $\Gamma_{\rm prep} = 0.10$, the R-direction response displays granular-like hardening ($\alpha \approx 2/3$), whereas the L-direction response shows biopolymer-like stiffening ($\alpha \approx 3/2$). 
Figure~\ref{fig:asymmetric}(c) reveals that the exponent $\alpha$ along the R- and L- directions can be tuned reciprocally between $2/3$ and $3/2$  by varying $\Gamma_{\rm prep}$ (top panel). Concurrently, a divergence emerges in the onset stiffening strain $\tau_c/G_0$ between R- and L- directions when $\Gamma_{\rm prep} \neq 0$ (bottom panel), confirming that this asymmetric stiffening results from the different distances to shear jamming boundaries in opposite directions. 

\subsection{Physical origin of $\alpha\approx3/2$ stiffening }

\begin{figure*}[t]
    \centering
    \includegraphics[width= \textwidth]{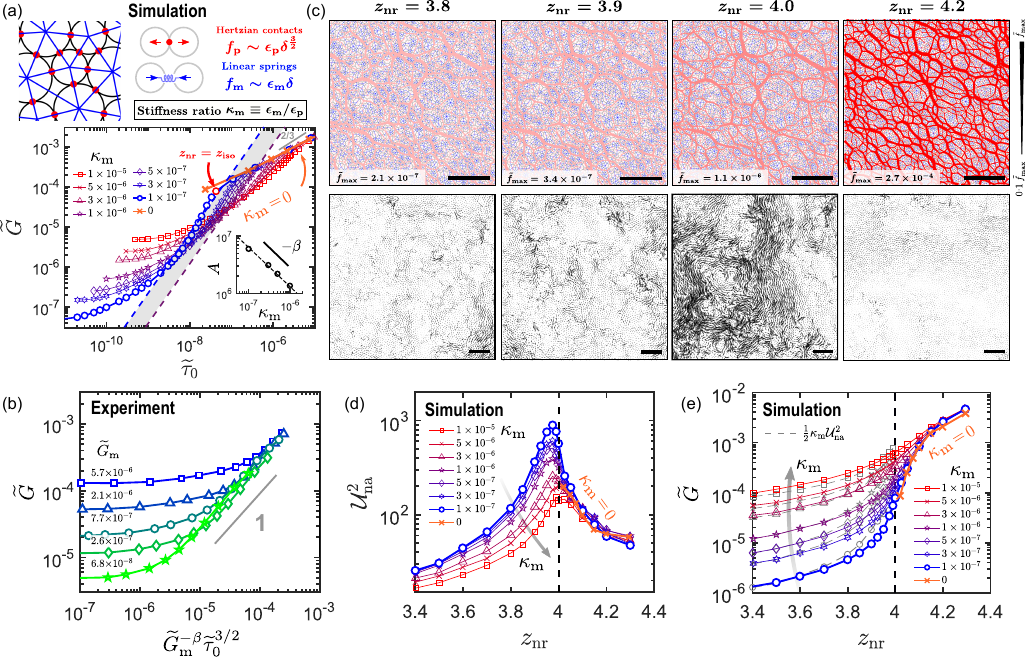}
    \caption{{\bf Physical origin of $\alpha\approx 3/2$ stiffening.}
    (a) Upper schematic: the coarse-grained particle-spring model in simulation. The blue lines and red dots represent inter-particle springs and contact points, respectively.
    Lower plot: Plots of the normalized shear modulus ($\widetilde{G}$) against the normalized shear stress ($\widetilde\tau_0$) for different stiffness ratio $\kappa_{\rm m}$
    (from simulations). Here, $\widetilde{G} = G/G_{\rm p}$ and $\widetilde{\tau_0} = \tau_0/G_{\rm p}$, where $G_{\rm p}$ is the effective shear modulus of the particles (Eq.~\ref{eq:method:Gp_sim}).
    The grey region indicates the scaling relation $\widetilde{G} = A \widetilde{\tau_0}^{3/2}$, where the prefactor 
    $A\sim\kappa_{\rm m}^{-\beta}$ with $\beta=0.68$ (see the inset). 
    The crossover from $\alpha = 3/2$ to $2/3$ occurs when the nearest neighbor number $z_{\rm nr}$ aligns with the isostatic point $z_{\rm iso} =4$. 
    (b) Plots of $\widetilde{G}$ against $\widetilde{G}^{-\beta}_{\rm m}\widetilde{\tau_0}^{3/2}$ (from experiments), where $\widetilde{G}$, $\widetilde{G}_{\rm m}$ and $\widetilde{\tau}_0$ are the rescaled parameters by normalizing the shear modulus of PS particles ($G_{\rm ps}= 1.6$~GPa). These dense soft composites were prepared with constant $\phi = 0.64$ and $\delta\gamma_{\rm osc} = 0.10$, but with different matrix moduli, where  $\widetilde{G}_{\rm m}= G_{\rm m}/G_{\rm p}$ ranged between $5.7\times 10^{-6}$ and $6.8 \times 10^{-8}$.
    (c) Simulation snapshots ($\kappa_{\rm m} = 10^{-7}$) show the contact force networks (top row) and corresponding particle displacements (bottom row) during strain stiffening when $z_{\rm nr}$ = 3.8, 3.9, 4.0 (={$z_{\rm iso}$}), and 4.2. The thickness of the red contact lines (top row) indicates the contact forces relative to the maximum contact force $\widetilde{f}_{\rm max}$ in each snapshot. The displacement arrows (bottom row) indicate the non-affine particle displacements ($\widetilde{\bf{u}}_{\rm na,i}$) within a step strain of $\delta\Gamma = 10^{-4}$, multiplied by $500$ for better visualization. 
    In simulations, $\widetilde{f}_{\rm max}$ is normalized by a characteristic contact force scale (Eq.~\ref{eq:method:force_unit}) and $\widetilde{\bf{u}}_{\rm na,i}$ has the unit of a particle diameter. Scale bars: ten particle diameters.
    (d, e) Plots of the non-affinity $\mathcal{U}^2_{\rm na}$ (Eq.~\ref{eq:main:non-affine})
    and $\widetilde{G}$ against $z_{\rm nr}$ for different $\kappa_{\rm m}$. The grey dashed line in panel (e) represents the shear modulus predicted by $\widetilde{G} = \kappa_{\rm m} \mathcal{U}^2_{\rm na}/2$. 
    All simulations were conducted using $\phi_{\rm eq} = 0.860$ and $\phi = 0.885$. 
    }
    \label{fig:simulation}
\end{figure*}

The stiffening exponent $\alpha \approx 3/2$ aligns with the values reported for biopolymer networks~\cite{gardel2004_science,lin2010_prl,Broedersz2014_RMP, Burla2019_NatPhys}, 
soft tissues~\cite{Song2025_np}, and biomimetic gels~\cite{Kouwer2013_nature,Jaspers2014_nc,Prince2024_bm}~(Extended Data Fig.~\ref{fig:extended:bio}). In those systems, such scaling is conventionally attributed to the entropic stiffening of semi-flexible polymers. However, nonlinear elasticity of the matrix cannot account for the mechanical responses observed in our PS-PDMS composites, as silicone gels exhibit a broad linear regime and only a weak stiffening ($\alpha <1$) at large strains (Extended Data Fig.~\ref{fig:extended:pdms}). Instead, the training-dependent stiffening behaviors (Fig.~\ref{fig:memory}) suggest structural memory as the key contributor to the $\alpha \approx 3/2$ exponent.

\begin{figure*}[]
    \centering
    \includegraphics[width=0.99 \textwidth]{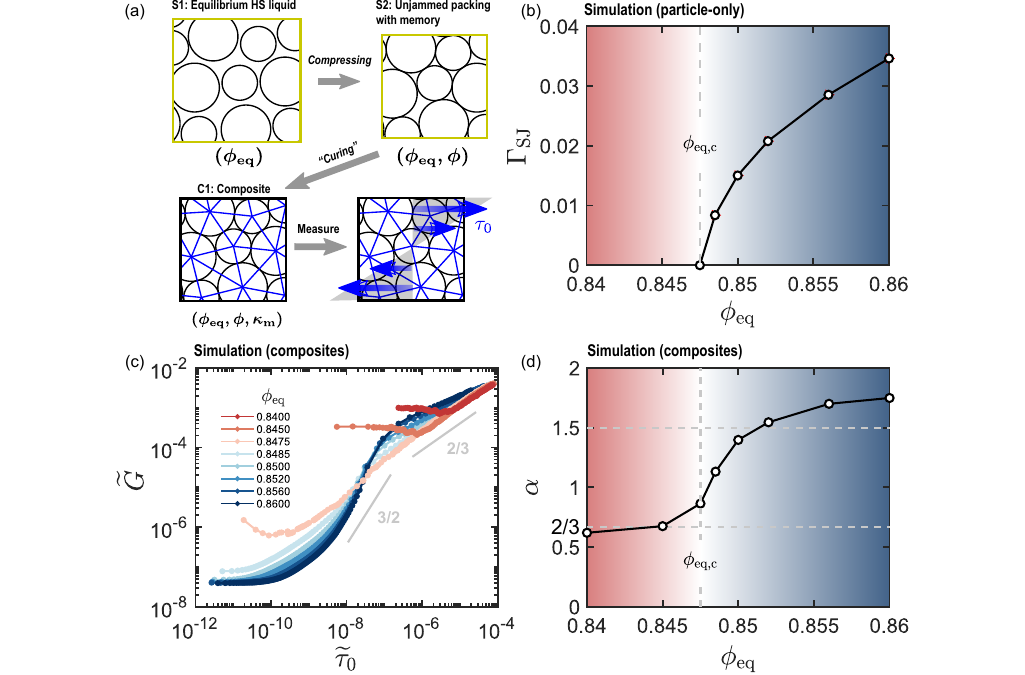}
    \caption{{\bf History-dependent shear jamming and strain stiffening in 2D simulations.}
    (a) Schematic illustrating the preparation of different particle configurations in the simulations. Distinct particle configurations are obtained by first preparing an equilibrium hard-sphere liquid at an initial packing fraction $\phi_{\rm eq}$ (S1) and then compressing the packing to reach a target packing fraction  $\phi$ (S2). 
    During the compression, the hard-sphere interaction between the particles is replaced by Hertzian interaction.
    Subsequently, linear springs are placed to connect the Voronoi neighbors. 
    (b) Shear jamming strain $\Gamma_{\rm SJ}$ for simulation systems prepared with different initial packing fractions ($\phi_{\rm eq}$) but the same final packing fraction ($\phi = 0.880$). (c) Plots of $\widetilde{G}$ against $\widetilde{\tau}_0$ for the systems with constant $\phi = 0.880$ and $\kappa_{\rm m} = 10^{-7}$, prepared with different $\phi_{\rm eq}$.  (d) Plot of $\alpha$ against $\phi_{\rm eq}$ for the stiffening curves shown in panel (c). Here, $\alpha$ denotes the largest stiffening exponent among the power-law exponents obtained by fitting $\widetilde{G}\propto \widetilde{\tau}_0^{\alpha}$ to the $\widetilde{G}(\widetilde{\tau}_0)$ results over intervals spanning a half decade of stress.
       }
    \label{fig:extend:alpha_phi_eq}
\end{figure*}

\begin{figure*}[]
    \centering
    \includegraphics[width=0.99 \textwidth]{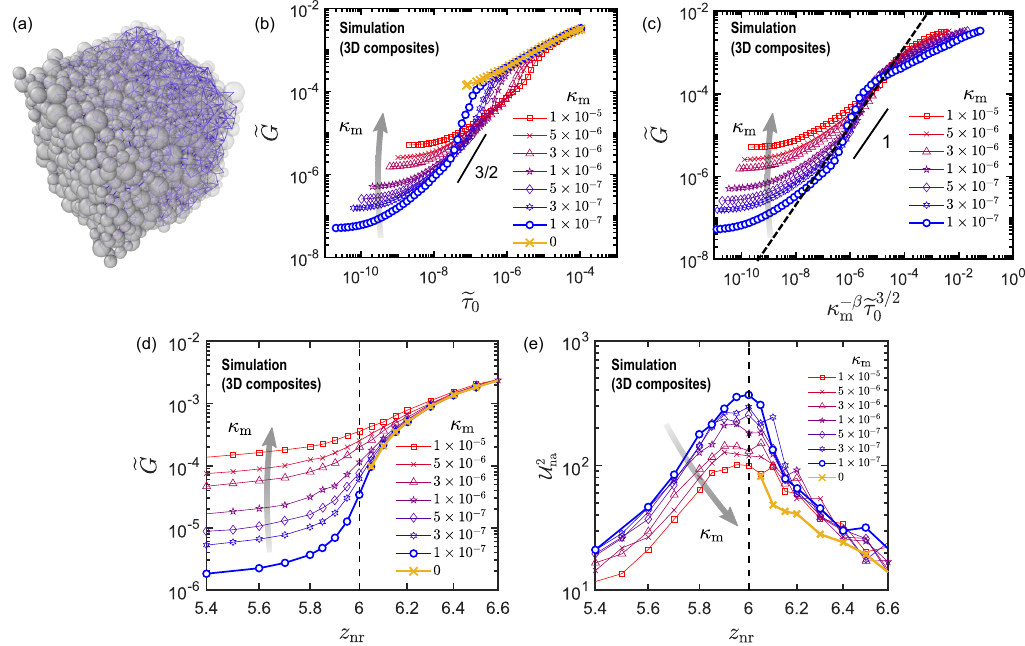}
    \caption{
    {\bf 3D simulations.}
    (a) Schematic of the 3D computational model. The blue bonds represent the linear springs between neighboring spheres. 
    (b) Plots of $\widetilde{G}$ against $\widetilde{\tau}_0$ with constant $\phi = 0.684$ and $\phi_{\rm eq} = 0.643$, while $\kappa_{\rm m}$ is varied between $1\times 10^{-5}$ and $1 \times 10^{-7}$. The $\kappa_{\rm m}=0$ results are obtained from particle-only simulations.
    (c, d) Plots of $\widetilde{G}$ against $\kappa_{\rm m}^{-\beta}\widetilde{\tau}_0^{3/2}$ 
    ($\beta = 0.68$)
    and $z_{\rm nr}$, respectively, using the same data as in (b).  
    (e) Plot of the non-affinity $\mathcal{U}^2_{\rm na}$ (Eq.~\ref{eq:main:non-affine}) against $z_{\rm nr}$.
    }
    \label{fig:extend:3d}
\end{figure*}

To accurately capture the interplay between the particle networks and the polymeric matrix, we developed a coarse-grained composite model consisting of dense Hertzian particles connected by linear springs (Fig.~\ref{fig:simulation}(a)). Particle configurations were generated by first preparing an equilibrium hard-sphere liquid state at a packing fraction $\phi_{\rm eq}$ and then quasi-statically compressing the system  to a target fraction $\phi > \phi_{\rm eq}$~\cite{Jin2021_pnas,pan2023_pnas_n}.
By systematically tuning $\phi_{\rm eq}$ while keeping $\phi$ fixed, this protocol yielded initial states with distinct memories:  $\Gamma_{\rm SJ}$ appeared to depend on $\phi_{\rm eq}$ above a critical preparation density $\phi_{\rm eq,c}$ (Extended Data Fig.~\ref{fig:extend:alpha_phi_eq}).
For a given particle configuration, a network of linear springs was placed between the centers of Voronoi-neighboring particles to capture the essential role of the polymeric matrix in a coarse-grained manner.
In simulations, the stiffness ratio of springs to particles ($\kappa_{\rm m}\equiv\epsilon_{\rm m}/\epsilon_{\rm p}$) was maintained from $10^{-7}$ to $10^{-5}$,  matching the modulus ratio between PS particles and PDMS gels in experiments. This particle-based simulation reproduced the crossover in the stiffening exponent ($\alpha$) with varying preparation history (Extended Data Fig.~\ref{fig:extend:alpha_phi_eq}). 
When $\phi_{\rm eq}\gg \phi_{\rm eq,c}$, the simulation obtained $\alpha \approx 3/2$ as the embedded particles were sufficiently annealed, analogous to the experimental regime of $\delta\gamma_{\rm osc} \gg \delta\gamma_c$. 
Since both two-dimensional (2D) and three-dimensional (3D) simulations produce quantitatively similar stiffening behaviors (Extended Data Fig.~\ref{fig:extend:3d}),
we herein elucidate the physical origin of $\alpha \approx 3/2$ using the 2D results, which allow for better visualization of the internal particle configurations and force networks.

We focus on the representative simulations with $\phi = 0.885$ and $\phi_{\rm eq} = 0.860$. Figure~\ref{fig:simulation}(a) plots the normalized shear modulus
$\widetilde{G} = G/G_{\rm p}$ against the normalized shear stress
$\widetilde{\tau}_0 = \tau_0/G_{\rm p}$ for different values of $\kappa_{\rm m}$, where $G_{\rm p}$ is the effective shear modulus of particles (see Eq.~\ref{eq:method:Gp_sim} in Method).
For $\kappa_{\rm m}$ between $10^{-7}$ and $10^{-6}$ (shaded gray region), the response follows biopolymer-like stiffening scalings $\widetilde{G} = A\widetilde{\tau_0}^\alpha$ with $\alpha\approx 3/2$. Beyond a critical stress, the stiffening transitions to a $\kappa_{\rm m}$-independent regime with $\alpha\approx 2/3$. The prefactor $A$ decrease with $\kappa_{\rm m}$ as a power law $A\sim\kappa_{\rm m}^{-\beta}$, where $\beta\approx 0.68$. For stiffer springs ($\kappa_{\rm m}= 3\times 10^{-6}, 5\times 10^{-6}$ and $10^{-5}$), the  stiffening response weakens markedly, consistent with the experimental results measured in stiffer matrix (Fig.~\ref{fig:memory}(f)). 

Given the analogy between $\kappa_{\rm m}\equiv\epsilon_{\rm m}/\epsilon_{\rm p}$ in simulations and $G_{\rm m}/G_{\rm p}$ in experiments, the stiffening scaling $\widetilde{G} = A\widetilde{\tau_0}^\alpha\sim \kappa_{\rm m}^{-\beta}{\tau_0}^{3/2}$ 
obtained from simulations implies a similar dependence on $G_{\rm m}/G_{\rm p}$ for the composite modulus ($\widetilde{G}$) measured in experiments.  To test this, the experimental results in Fig.~\ref{fig:memory}(f) were re-plotted as $\widetilde{G}$ versus $\widetilde{G}_{\rm m}^{-\beta}\widetilde{\tau_0}^{3/2}$, where all parameters are normalized by the modulus of PS particles $G_{\rm ps} = 1.6$~GPa. As shown in Fig.~\ref{fig:simulation}(b), the stiffening regimes for various $G_{\rm m}$ align with a single, $G_{\rm m}$-independent master curve, $\widetilde{G}\sim  \widetilde{G}_{\rm m}^{-\beta}\widetilde{\tau_0}^{3/2}$, confirming the validity of the simulation prediction. 

We further analyzed the evolution of the contact network and non-affine deformation as a function of the average non-rattler contact number ($z_{\rm nr}$) using the simulation data for $\kappa_{\rm m}= 10^{-7}$ (see Fig.~\ref{fig:simulation}(c)). 
Applied shear drives a monotonic growth in $z_{\rm nr}$ near isostaticity $z_{\rm iso}=4$ (top row), while the particle networks remain sub-isostatic ($z_{\rm nr}<z_{\rm iso}$) throughout the $3/2$-stiffening regime. 
Conversely, non-affine particle displacements peak at $z_{\rm nr} = z_{\rm iso}$ but decay both below and above this threshold (bottom row).  
The non-affinity was quantitatively expressed as~\cite{Broedersz2014_RMP,Sharma2016_NatPhys}:
\begin{equation}\label{eq:main:non-affine}
    \mathcal{U}_{\rm na}^2 = \frac{1}{N\delta\Gamma^2}\sum_{i=1}^N|\widetilde{\mathbf{u}}_{{\rm na},i}|^2,
\end{equation}
where $N$ is the number of particles, $\delta\Gamma = 10^{-4}$ is the strain increment, and $\widetilde{\mathbf{u}}_{\rm na,i}$ is the non-affine displacement of particle $\it i$ normalized by the mean particle diameter. 
Across all $\kappa_{\rm m}$ values, $\mathcal{U}_{\rm na}^2$ attains a pronounced maximum near $z_{\rm iso}$ (Fig.~\ref{fig:simulation}(d)). Above isostaticity ($z_{\rm nr}>z_{\rm iso}$), the divergence of non-affinity ($\mathcal{U}_{\rm na}^2$) as $z_{\rm nr}\rightarrow z_{\rm iso}^+$ stems from the marginal stability of jammed packings~\cite{Ellenbroek2006_prl,Wyart2008_prl}.
Below isostaticity ($z_{\rm nr}<z_{\rm iso}$), the rise in $\mathcal{U}_{\rm na}^2$ as $z_{\rm nr}\rightarrow z_{\rm iso}^-$ is reminiscent of the behavior of sub-isostatic networks with separated stiffness components~\cite{Wyart2008_prl}.

To show the synergistic role of particle contacts and non-affine displacements in governing the stiffening transition, we re-plot $\widetilde{G}$ versus $z_{\rm nr}$ for different $\kappa_{\rm m}$ in Fig.~\ref{fig:simulation}(e). For $z_{\rm nr} > z_{\rm iso}$, $\widetilde{G}(z_{\rm nr})$ collapses onto the simulation prediction for $\kappa_{\rm m}=0$, confirming that the exponent $\alpha \approx 2/3$ originates from the shear hardening of granular packings~\cite{pan2023_pnas_n}. 
For  $z_{\rm nr}< z_{\rm iso}$, the shear modulus can be well approximated by  
$\widetilde{G} \sim \kappa_{\rm m} \mathcal{U}^2_{\rm na}$ (gray dashed lines), as derived from  the linear theory of disordered networks (Figs.~S3 and S4). Hence, the stiffening exponent $\alpha \approx 3/2$ observed below isostaticity  arises from an enhanced elastic energy stored in the matrix, driven by amplified non-affine particle rearrangements near $ z_{\rm iso}$. 



\begin{figure*}[t]
    \centering
    \includegraphics[width = \textwidth]{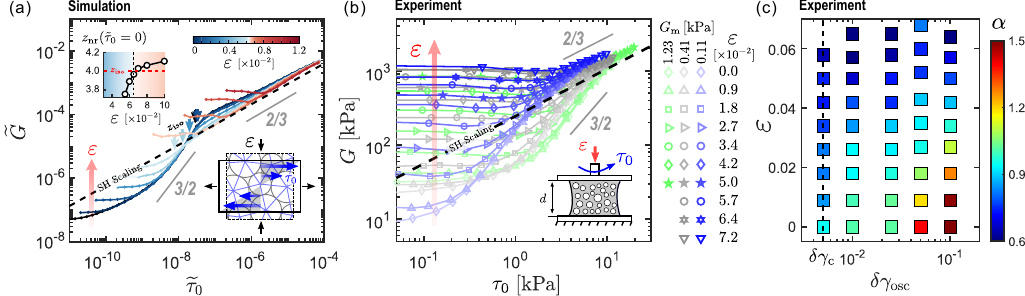}
    \caption{
    {\bf Programmable strain-stiffening under multi-axial strain.} 
    (a) Plots of $\widetilde{G}$ versus $\widetilde{\tau}_0$ under different axial strains ($\varepsilon$) (from simulations). The simulations were conducted with constant system parameters ($\phi_{eq} = 0.860$, $\phi = 0.885$, and $\kappa_{\rm m}= 10^{-7}$) and with $\varepsilon$ ranging between 0 and $0.015$. The inset plot represents the zero-stress coordination number $z_{\rm nr} (\widetilde{\tau}_0=0)$ as the function of $\varepsilon$, where this value exceeds the iso-static point $z_{\rm iso}=4$ when $\varepsilon>0.0065$. For the stiffening curves with $z_{\rm nr}(\widetilde{\tau}_0=0)<z_{\rm iso}$, the $\widetilde{\tau}_0$ value above which the shear-induced $z_{\rm nr}$ exceeds $z_{\rm iso}$ are marked by the downward arrows with the same color.
    (b) Plots of $G$ against $\tau_0$ under different $\varepsilon$ for three distinct composites, prepared with the constant particle volume fraction $\phi = 0.64$ and training amplitude $\delta\gamma_{\rm osc} = 0.1$ but different matrix moduli ($G_{\rm m} = 0.11$, 0.41, and 1.23~kPa). The dashed lines in (a, b) indicate the shear-hardening scaling predicted by Eq.~\ref{eq:sh}. 
    (c) Colormap plot of the maximum stiffening exponent $\alpha$ versus $\delta\gamma_{\rm osc}$ and $\varepsilon$ for soft composites with constant $\phi = 0.64$ and $G_{\rm m} = 0.11$ kPa.}
    \label{fig:multiaxial}
\end{figure*}

In contrast to conventional mechanisms based on the nonlinear elasticity of polymers~\cite{meng2016_soft, van_Oosten2019_Nature, Song2025_np}, our findings reveal a distinct path to biopolymer-like responses: one that relies exclusively on the structural characteristics of disordered systems near rigidity transitions~\cite{Wyart2008_prl, Broedersz2011_NatPhys,shivers2025_prx}. To assess the generality of this mechanism,  we further demonstrate that the same $\alpha\approx 3/2$ stiffening emerges in dense composites of varying material compositions, including solid glass (GS)-PDMS and hollow glass sphere (HGS)-hydrogel composites (Extended Data Fig.~\ref{fig:extended:bio}).

\noindent
\subsection{\bf {Multi-axial programmability} }

Beyond simple shear, strain-stiffening under orthogonal loading is another essential feature to be engineered for many biomimetic materials~\cite{van_Oosten2019_Nature,Shivers2020_PNAS,song2021_jap,Xie2021_NatCom}. We herein demonstrate systematic control over the stiffening exponent in dense soft composites through the application of multi-axial strain. 

In simulations, we subjected a particle-spring network (with $\phi_{\rm eq} = 0.860$, $\phi = 0.885$, and $\kappa_{\rm m}= 10^{-7}$) to a pure shear (characterized by axial strain $\varepsilon$) followed by a superimposed simple shear (characterized by shear stress $\widetilde{\tau}_0$). Figure~\ref{fig:multiaxial}(a) shows the simulated $\widetilde{G}$ as a function $\widetilde{\tau}_0$ for axial strains ranging from $\varepsilon=0$ to 1.51~\%. Under small compression ($\varepsilon <0.65$~\%), the system initially resides below isostaticity ($z_{\rm nr}<z_{\rm iso} = 4$) at zero-stress $\widetilde{\tau}_0 = 0$ (inset). As $\varepsilon$ increases and $z_{\rm nr}$ passes beyond $z_{\rm iso}$, the maximum stiffening exponent $\alpha$ transitions from $3/2$ to $2/3$. For $\varepsilon > 0.65$~\%, the axial compression induces a pre-jammed state ($z_{\rm nr} > z_{\rm iso}$) at zero-stress, and the subsequent shear yields only the granular-like stiffening response ($\alpha\approx 2/3$).

This multi-axial control was further validated experimentally. Soft composites prepared with the constant $\phi =0.64$ and $\delta\gamma_{\rm osc}=0.1$, but varying matrix stiffnesses ($G_{\rm m}= 0.11, 0.41, 1.23$~kPa), exhibit analogous behavior (Fig.~\ref{fig:multiaxial}(b)): with increasing axial compression, the biopolymer-like ($\alpha \approx 3/2$) regime vanishes, converging to a matrix-independent granular scaling ($\alpha \approx 2/3$). Thus, in addition to $\delta\gamma_{\rm osc}$, the axial strain ($\varepsilon$)  serves as a key control parameter for tuning the proximity of the particle configuration to the shear-jamming boundary (Fig.~S5).  Figure~\ref{fig:multiaxial}(c) maps the maximum stiffening exponent $\alpha$ (spanning from $2/3$ to $3/2$) as a function of both $\varepsilon$ and $\delta\gamma_{\rm osc}$ for $G_{\rm m}= 0.11$~kPa.

\section{Conclusions}

This study presents a design strategy to control the strain-stiffening behavior of dense soft composites by manipulating the structural memory of inclusions.  We implement a training protocol that adjusts particle networks through a history-dependent shear-jamming transition  (Fig.~\ref{fig:training}), enabling a crossover from granular-like ($\alpha \approx 2/3$) to biopolymer-like ($\alpha \approx 3/2$) stiffening responses (Figs.~\ref{fig:memory} and \ref{fig:asymmetric}).  When the embedded particles are pre-jammed, the stiffening exponent $\alpha \approx 2/3$ aligns with granular shear-hardening scaling (Eq.~\ref{eq:sh}). Conversely, when the trained particle configurations remain below but close to a rigidity transition,  we observed an enhanced stiffening exponent $\alpha \approx 3/2$, induced by increased elastic energy from significant non-affine particle rearrangements close to jamming  (Fig.~\ref{fig:simulation}). Our coarse-grained simulations identify the isostatic point ($z_{\rm iso}$) as the critical threshold that governs the transition between these two stiffening regimes. Notably,  we predict a continuous tuning of $\alpha$ from $2/3$ to $3/2$ through a combination of simple and pure shear, a finding that is validated experimentally in composites subjected to multi-axial strain (Fig.~\ref{fig:multiaxial}). 

From the perspectives of soft material engineering, an enhanced stiffening exponent $\alpha \approx 3/2$ is a signature of tissue-like or biopolymer-like mechanics, typically achieved through the nonlinear elasticity of polymer matrices~\cite{Kouwer2013_nature,Prince2024_bm,Song2025_np}. 
In contrast, our experiments and simulations uncover an alternative mechanism for designing biomimetic materials, relying solely on the non-equilibrium characteristics of disordered contact networks. 
Given the prevalence of structural memory in granular systems~\cite{Kumar2016_gm,Jin2021_pnas,kawasaki2024_prl}, biopolymer-like stiffening responses can be achieved through mechanical pre-training with diverse material compositions.

\bibliography{main.bib}

\clearpage
\noindent 
{\large \bf Method}

\vspace{11pt}

\beginSI

\noindent {\bf Materials}

\begin{figure*}[!t]
    \centering
    \includegraphics[width=0.99 \textwidth]{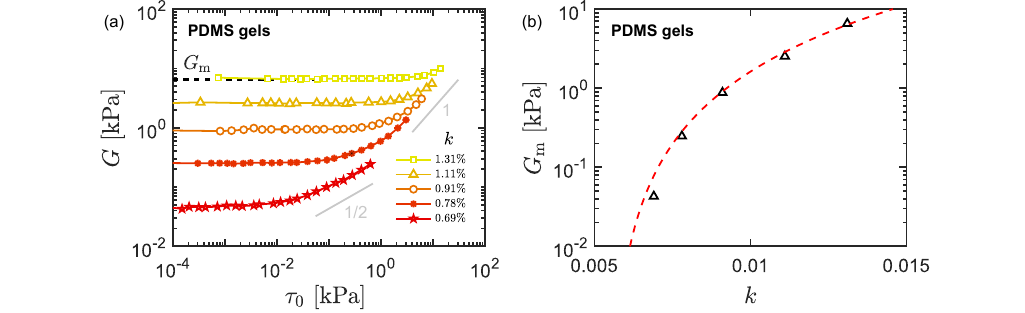}
    \caption{{\bf Nonlinear elasticity of PDMS gels.}
        (a) Plot of  shear modulus ($G$) against shear stress ($\tau_0$) for PDMS gels with varying crosslink density $k$.
        For each $k$, the linear modulus ($G_{\rm m}$) denotes $G$ measured at small $\tau_0$. 
        (b) Plot of $G_{\rm m}$ against $k$ for the PDMS gels. The red dashed curve indicates a best-fit power-law relation $G_{\rm m} = A(k-k_{\rm c})^{\beta}$ with $A=2.05$ GPa, $k_{\rm c} = 0.0055$, and $\beta = 2.6$.
    }
    \label{fig:extended:pdms}
\end{figure*}

\noindent 
{\it The PDMS-based soft composites} ---
Our PDMS-based soft composites consist of polystyrene (PS) microspheres (XMO-50, Dongguan Xinmiao New Material Co.) with a mean diameter of 29~$\mu$m randomly embedded in a crosslinked polydimethylsiloxane (PDMS) gel matrix. To fabricate composites with structural memory, we applied oscillatory training to the suspension consisting of PS particles and the uncrosslinked PDMS melts. After training, the rheometer plates were maintained still for three hours to allow the precursor slowly cure into a soft gel. The precursor contains a silicone base (DMS-V31, Gelest Inc.), crosslinkers (HMS-301, Gelest Inc.), and
a catalyst (SIP6831.2, Gelest Inc.). We used a catalyst weight ratio of 0.015\% to ensure that PDMS remained fluid during training and that the crosslinking process was completed within three hours.
The elasticity of the matrix was controlled by the crosslink density $k$~\cite{Zhao2022_SoftMatter}. We tuned $k$ between 0.69 \% and 1.33 \%, yielding PDMS gels whose linear shear modulus $G_{\rm m}$ ranges from 0.1 kPa to 10 kPa (Extended Data Fig.~\ref{fig:extended:pdms}).

\noindent
{\it The hydrogel-based soft composites} ---
The matrix of the hydrogel composites is a crosslinked polyacrylamide (PAAm) network swollen in an aqueous sodium alginate solution.
The PAAm hydrogel was fabricated by sequentially dissolving sodium alginate (Macklin Biochemical Technology), acrylamide (AAm from  Sigma-Aldrich), crosslinker N,N'-methylenebisacrylamide (MBAA from Sigma-Aldrich), initiator ammonium persulfate (APS from Sigma-Aldrich), and accelerator N,N,N',N'-tetramethylethylenediamine (TEMED from Sigma-Aldrich) in deionized water. The MBAA-to-AAm ratio was varied to tune the elasticity of the PAAm hydrogel. The weight percentage of other components was kept as follows: sodium alginate (0.75 wt \%), AAm (5.96 wt \%), MBAA (0.01 wt \%), APS (0.19 wt \%), and TEMED (0.02 wt \%). To fabricate hydrogel composites, we applied oscillatory shear training to the mixture of the particles and the precursor of the PAAm gel, and then let the gel crosslink for ten hours before measurements. 

\vspace{11pt}

\noindent {\bf Superposition rheology}

\noindent 
The shear moduli of soft composites were measured using a commercial rheometer (Anton Paar MCR302) equipped with a $25$-mm parallel-plate shear cell. The shear strain was calculated from the angular displacement $\theta$ of the top plate as $\Gamma \equiv \frac{2}{3} R\theta/d$, where $R$ is the radius of the measuring plate and $d$ is the distance between the two parallel plates. For each experiment, we increased the imposed strain stepwise. At each pre-strain $\Gamma_0$, we held the plates still for two minutes to allow sample to relax. 
Then, we superposed a small oscillatory shear strain on top of $\Gamma_0$: $\Gamma(t) = \Gamma_0 + \delta\Gamma \sin \omega t$ with $\delta\Gamma = 0.1\%$ and $\omega = 1.27$ rad/s for four cycles (Fig.~S1). 
We measured the resultant evolution of shear stress $\tau$, which was converted from the torque $T$ measured from the rheometer following $\tau\equiv \frac{4}{3}\frac{T}{\pi R^3}$. 
Data from the last two cycles were fitted to
$\tau(t) = \tau_0 + \delta\tau \sin(\omega t + \Delta)$. 
The storage moduli and the loss moduli were calculated as $G^{\prime}\equiv \frac{\delta\tau}{\delta\gamma} \cos \Delta$ and $G^{\prime\prime}\equiv \frac{\delta\tau}{\delta\gamma} \sin \Delta$. 
The traces of $G(\tau_0)$ gradually varied with the maximum pre-strain $\Gamma_{\rm 0,max}$ due to the Mullins effect~(see Fig.~S6 and Refs. \cite{mullins1969_RCT, Song2025_np}), while the stiffening exponent $\alpha$ remains unchanged. 
In the main text, we fixed $\Gamma_{\rm 0, max} = 0.19$ for composites prepared with $\delta\gamma_{\rm osc} >  \delta\gamma_{\rm c} \approx 0.005$, which allowed measuring a wide-range of nonlinear responses. For composites prepared with $\delta\gamma_{\rm osc} =  \delta\gamma_{\rm c}$ a smaller $\Gamma_{\rm 0, max} = 0.05$ was used in the measuring the stiffening curves in Fig.~\ref{fig:memory}(e) to avoid wall slip.

\vspace{11pt}

\noindent {\bf Traction force microscopy}

\begin{figure*}[!t]
    \centering
    \includegraphics[width=0.99\textwidth]{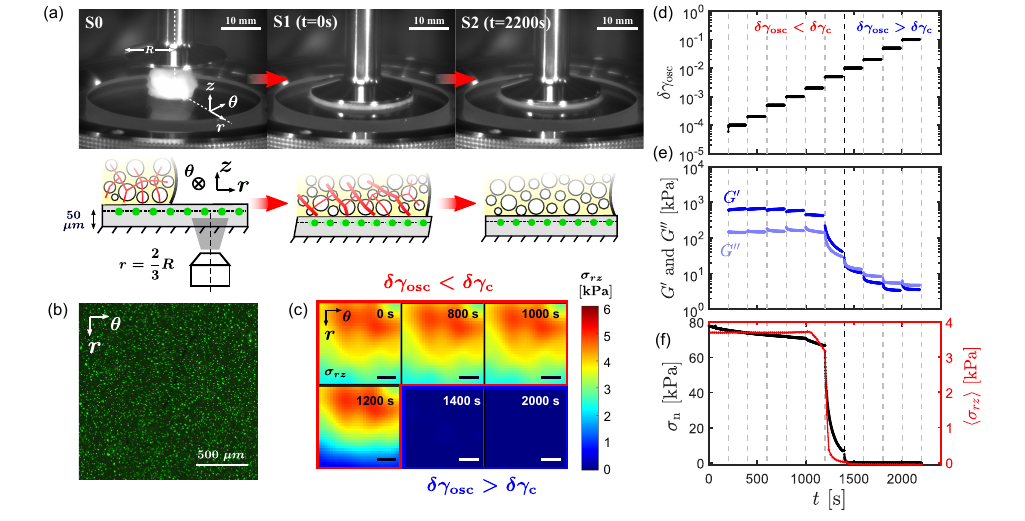}
    \caption{{\bf Oscillatory training of PS-PDMS suspensions.}
    (a) Top panels: First, a  PS-PDMS suspension with $\phi = 0.64$ was placed between a shear plate and a glass substrate (S0 state). Second, the top plate compressed the sample to achieve a gap distance $ h = 1$ mm (S1 state). Finally, the suspension was trained by an oscillatory shear ($\delta\gamma_{\rm osc}$) (S2 state). Bottom panels: Schematics of traction force microscopy (TFM) used to measure local stresses. (b) Snapshot of fluorescent tracer particles at the suspension-elastomer interface. 
    (c) Snapshots of the boundary stress map ($\sigma_{rz}$) measured at various times. Scale bars: 500 $\mu$m.
    (d) Stepwise increase of $\delta\gamma_{\rm osc}$ with time ($t$) imposed on an as-prepared PS-PDMS suspension with $\phi = 0.64$. The grey shaded area highlights the unjamming threshold $\delta\gamma_{\rm osc} = \delta\gamma_{\rm c} = 0.005$. Each $\delta\gamma_{\rm osc}$ was applied to the suspension samples for three minutes with the same angular frequency $\omega = 100$ rad/s. 
    (e) Plot of the complex moduli ($G^{\prime}$ and $G^{\prime\prime}$) versus $t$. The storage modulus $G^{\prime}$ falls below the loss modulus $G^{\prime\prime}$ when $\delta\gamma_{\rm osc}\geq \delta\gamma_{\rm c}$, indicating a complete relaxation of jammed contact networks.
    (f) Plot of the normal stress ($\sigma_{\rm n}$) and the averaged local radial stress ($\langle \sigma_{rz}\rangle$) against $t$, respectively. With the increase in $\delta\gamma_{\rm osc}$, both $\sigma_{\rm n}$ and $\langle\sigma_{rz}\rangle$ progressively decrease and vanish for $\delta\gamma_{\rm osc}\geq\delta\gamma_{\rm c}$.
            }
    \label{fig:extend:unjamming}
\end{figure*}

\noindent 
We used traction force microscopy (TFM)~\cite{style2014_sm} to measure the local in-plane stresses at the interface between  suspension samples and the substrate (Extended Data Fig.~\ref{fig:extend:unjamming}). To perform these measurements, the glass substrate of the shear cell was spin-coated with a thin layer of PDMS elastomer that had a thickness of 50~$\mu$m and a Young's modulus of 6.6~kPa. We then placed 5~$\mu$m-sized fluorescent beads on the elastomer surface as tracers. To prevent the shear-induced slipping of these beads, an additional 6-$\mu$m-thick PDMS layer was added on top of the first layer. By tracking the bead displacements using an objective placed at $2R/3$ radial position, the in-plane stress map in a 2.1~mm $\times$ 2.1~mm window was quantified based on the linear elasticity of the elastomer film~\cite{hu2024_jor}. By tracking the displacements of the 5~$\mu$m-sized fluorescent beads deposited on the elastomer surface, the in-plane deformations at the suspension-substrate interface were measured {\em in-situ}. Extended Data Fig.~\ref{fig:extend:unjamming} reports the evolution of stress component $\sigma_{\rm rz}$ during the oscillatory training.

\vspace{11pt}

\noindent {\bf{X-ray micro-computed tomography}}

\begin{figure*}[!t]
    \centering
    \includegraphics[width=0.99 \textwidth]{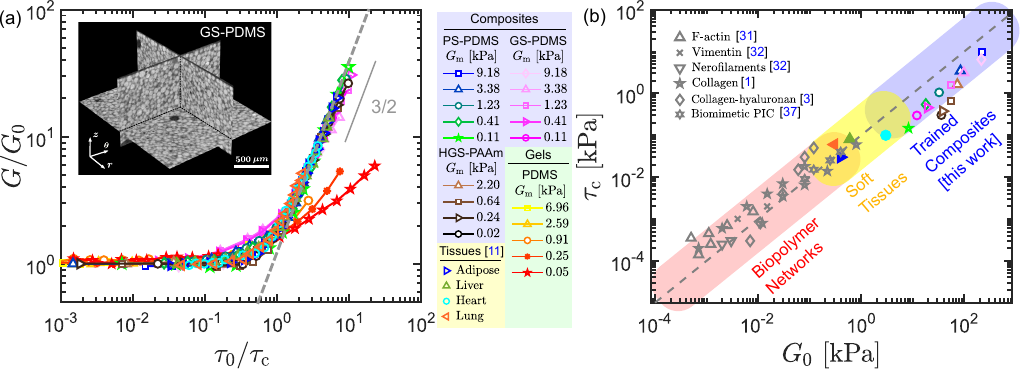}
    \caption{{\bf Strain stiffening in trained soft composites, soft tissues, and biopolymer networks.}
        (a) Plots of the normalized shear modulus ($G/G_0$) against the normalized shear stress ($\tau_0/\tau_c$) for different soft materials, where $G_0$ represents the shear modulus in the linear regime at low stresses, and $\tau_c$ marks the onset of strain-stiffening. The plot includes the data from: (i) PS-PDMS composites, adapted from Fig.~\ref{fig:memory}(f); (ii) GS-PDMS composite ($\phi = 0.64, \delta\gamma_{\rm osc} = 0.10$) prepared with different $G_{\rm m}$;  (iii) HGS-PAAm composites ($\phi = 0.62, \delta\gamma_{\rm osc} = 0.10$) prepared with different $G_{\rm m}$; (iv) soft PDMS gels with different $G_{\rm m}$; and (v) various soft tissues, adapted from Ref.~\cite{Song2025_np}. Inset: Micro-CT image revealing the microstructure of particle packing in a glass-in-PDMS composite ($\phi=0.64$).
        (b) 
        Plots of $\tau_c$ against $G_0$ for soft biopolymer networks, soft tissues, and the trained soft composites. The data of bio-polymer networks are adapted from Ref.~\cite{gardel2004_prl} (F-actin), Ref.~\cite{lin2010_prl} (Vimentin and Nerofilaments), Ref.~\cite{Licup2015_pnas} (Collagen), Ref.~\cite{Burla2019_NatPhys} (Collagen-hyaluronan composite networks), and Ref.~\cite{Kouwer2013_nature} (Biomimetic PIC networks); the data of soft tissues are adapted from Ref.~\cite{Song2025_np}; the data for soft composites  are taken from the panel (a).  The gray dashed line shows $\tau_{\rm c}= \Gamma_{\rm c}G_0$ with $\Gamma_{\rm c}=0.1$.
          }
    \label{fig:extended:bio}
\end{figure*}

\noindent 
The packing structures of glass spheres in a PDMS matrix shown in Extended Data Fig.~\ref{fig:extended:bio} were visualized via X-ray micro-computed tomography ($\mu$-CT). The X-ray experiments were conducted for a 3mm x 3mm x 1mm sample in a Cougar EVO X-ray inspection system (Comet Yxlon). The X-ray energy level was set to 66 keV. Tomography data were reconstructed with VGSTUDIO, achieving a spatial resolution of 1.5 $\mu$m.

\vspace{11pt}

\noindent {\bf Composite Simulation}

\noindent 
The composite consists of a disordered assembly of particles connected through linear elastic springs. The total potential energy is the sum over all $i, j$ of the pair interaction potentials: 
\begin{equation}\label{eq:method:total_potential}
V = \sum_{i=1}^N\sum_{j=1}^N \Bigl(V_{{\rm m}, ij}(r_{ij}) +  V_{{\rm p},ij}(r_{ij})\Bigr),
\end{equation}
where $V_{{\rm p},ij}$ and $V_{{\rm m},ij}$ are the pair interaction potentials for contact forces and matrix springs, respectively. The potential for contact forces between the $i$th and the $j$th particles is defined as
\begin{equation}\label{eq:method:hertz}
V_{{\rm p},ij}(r_{ij})  = \frac{2}{5} {\epsilon_{\rm p}} {\left(1-\frac{r_{ij}}{D_{ ij}}\right)}^{\frac{5}{2}}{H\left({1-\frac{r_{ij}}{D_{ ij}}}\right)},
\end{equation}
where the pre-factor $\epsilon_{\rm p}$ characterizes the strength of the interaction, $r_{ij}$ is the center-to-center distance, $D_{ij} = (D_i + D_j)/2$ {is} the average diameter, and $H(x)$ is the Heaviside step function. 
The pair interaction potential for a matrix spring connecting $i$th and $j$th particles that are originally Voronoi neighbours is
\begin{equation}\label{eq:method:spring}
    V_{{\rm m},ij}(r_{ij}) = \frac{1}{2}{\epsilon_{\rm m}}\left(1 - \frac{r_{ij}}{l_{ij}}\right)^2,
\end{equation}
where $l_{ij}$ is the original length of the spring, which is set by the initial center distance between the two particles at the curing stage.

All reported results were obtained from simulations containing $N = 2000$ grains in 3D (or $N = 16000$ in 2D), averaged over at least 64 independent samples. The particle diameters $D$ follows an inverse power-law distribution, $P(D) \sim D^{-d},~D_{\rm min} \le D \le D_{\rm min}/0.45$ (where $d$ is space dimension).
All deformations were applied quasi-statically, {\it i.e.}, via an affine deformation followed by an energy minimization. Specifically, shear stiffening was measured under constant-volume, simple shear in the x-z plane. The FIRE algorithm was used to minimize potential energy \cite{bitzek2006structur}, and the algorithm was terminated if the averaged residual force magnitude was smaller than $5 \times 10^{-13}$. After energy minimization, the stress tensor was calculated following the virial formula 
\begin{equation}\label{eq:method:stress_tensor}
\overline{\overline{\sigma}} = \frac{1}{L^{d}}\sum_{ij}{\bf r}_{ij}\otimes f_{ij},
\end{equation}
where ${\bf r}_{ij}$ is the branch vector connecting the centers between $i$th and $j$th particles, ${\bf f}_{ij}$ is the force, $L$ is the size of the simulation cell, and $d$ is the spatial dimension. We denote the shear stress as
\begin{equation}
    \tau\equiv -\overline{\overline{\sigma}}_{xz}
\end{equation} 
and the pressure as
\begin{equation}\label{eq:method:p}
p\equiv \frac{1}{d} Tr(\overline{\overline{\sigma}}).
\end{equation}
We also calculated the non-rattler contact number
\begin{equation}\label{eq:method:z_nr}
    z_{\rm nr} = \frac{1}{N_{\rm nr}}\sum_{i = 1}^{N}z_i\Theta_i,
\end{equation}
where $N_{\rm nr}$ is the total number of non-rattler particles, $N$ is the total number of particles, $z_i$ is the number of contacts for the $i$th particle, $\Theta_i=1$ (or 0) if the $i$th particle was (or not) a non-rattler particle. A non-rattler particle is defined as a particle with at least $d+1$ contacts, where $d$ is the spatial dimension. 
The shear modulus $G$ is calculated by $G=\partial \tau/\partial \Gamma$, where $\Gamma$ is the simple shear strain. 
{\color{orange}}

\noindent
{\it The normalized modulus and contact forces} ---
The units for length ($[\mathcal{L}]$), mass ($[\mathcal{M}]$), and energy ($[\mathcal{E}]$) were set to the mean grain diameter {$\langle D\rangle$}, mean grain mass and the strength of the Hertzian contacts, $\epsilon_{\rm p}$, from Eq.~\ref{eq:method:hertz}, respectively. To define a particle shear modulus $G_{\rm p}$ for the purpose of normalizing $G$ and $\tau_0$, we considered the contact force between two identical spheres with a mean diameter $\langle D\rangle$, 
\begin{equation}\label{eq:method:force_sim}
f = \epsilon_{\rm p}\langle D\rangle^{-1}\delta ^{\frac{3}{2}},
\end{equation}
where $\delta \equiv 1 - r/\langle D\rangle $, and $r$ was the center-to-center distance between the two spheres. This equation aligns with the Hertzian contact force law:
\begin{equation}\label{eq:method:force_exp}
    f =\frac{2}{3}E^*_{\rm p}\langle D\rangle ^{2}\delta^{\frac{3}{2}},
\end{equation}
where $E^*_{\rm p}\equiv E_{\rm p}/2(1-\nu^2_{\rm p})$, and $E_{\rm p}$ and $\nu_{\rm p}$ are the Young's modulus and the Poisson's ratio of the particle. Comparing Eq.~\ref{eq:method:force_sim} with Eq.~\ref{eq:method:force_exp}, we obtained a relation between the energy unit $\epsilon_{\rm p}$ and the reduced particle modulus $E_{\rm p}^*$:
\begin{equation}\label{eq:method:epsilon_p}
    \epsilon_{\rm p} = \frac{2}{3}E^*_{\rm p}\langle D\rangle^3.
\end{equation}
Considering $G_{\rm p} = E_{\rm p}/2(1+\nu_{\rm p})$ for isotropic materials, we obtained that
\begin{equation}\label{eq:method:Gp_sim}
    G_{\rm p} = \frac{3}{2}(1-\nu_{\rm p})\epsilon_{\rm p}\langle D\rangle ^{-3}.
\end{equation}
In Figs.~\ref{fig:simulation} and~\ref{fig:multiaxial}, we used $\nu_{\rm p}=\nu_{\rm PS} = 0.35$ for the PS particles~\cite{Ryusuke1960_JPSJ},  $\epsilon_{\rm p}=1$, and $\langle D\rangle=1$ to calculate $G_{\rm p}$.
 From Eqs.~\ref{eq:method:force_sim} and~\ref{eq:method:epsilon_p}, the force unit  ($[\mathcal{F}]$) was defined as
\begin{equation}\label{eq:method:force_unit}
    [\mathcal{F}]\equiv[\mathcal{E}][\mathcal{L}]^{-1}=\epsilon_{\rm p}\langle D\rangle^{-1}=\frac{2}{3}E_{\rm p}^*\langle D\rangle^2.
\end{equation}

\noindent 
{\it The non-affine displacements} ---
To compute the non-affine response of a composite system under a shear strain $\Gamma_0$, we imposed a cyclic shear with a strain step $\Delta\Gamma = 10^{-5}$, while maintaining the strain amplitude as small as $10^{-4}$. When the mechanical response became reversible, the non-affine displacements were calculated as
\begin{equation}\label{eq:method:u_i}
 {\bf u}_{{\rm na},i}(\Gamma_0) \equiv \mathbf{r}_i(\Gamma_0+\Delta\Gamma)-
 {\overline{\overline{F}}}
 \mathbf{r}_i(\Gamma_0),   
\end{equation}
where $\mathbf{r}_i$ is the position of the $i$th particle and 
{$\overline{\overline{F}}$}
is the affine deformation gradient. For each shear cycle, we computed the average $\mathbf{u}_{{\rm na},i}$ measured at five representative pre-strains: $\Gamma_0-10\Delta\Gamma$, $\Gamma_0-5\Delta\Gamma$, $\Gamma_0$, $\Gamma_0+5\Delta\Gamma$, and $\Gamma_0+10\Delta\Gamma$. The normalized non-affine displacement 
\begin{equation}
\widetilde{\mathbf{u}}_{{\rm na},i} = \mathbf{u}_{{\rm na},i}/ \langle D\rangle,
\end{equation}
where $\langle D \rangle$ is the mean particle diameter. 

\vspace{11pt}

\noindent {\large \bf Data availability}

\noindent
Additional data regarding the study are available from the corresponding authors upon request. 

\vspace{11pt}

\noindent {\large \bf Acknowledgments}

\noindent
This work was supported by the General Research Fund (No.~16307422 and No.~16306723) and the Collaborative Research Fund (No.~C6004-22Y and No.~C6041-24G-B) from the Hong Kong Research Grants Council (RGC). Y.Z. acknowledges the funding support from the RGC Postdoctoral Fellowship (No. PDFS2324-6S02). D.P. acknowledges  funding from the National Natural Science Foundation of China (No.~12404290). We acknowledge the use of the High Performance Cluster at Institute of Theoretical Physics, Chinese Academy of Sciences.

\vspace{11pt}

\noindent {\large \bf Author contributions}

\noindent 
Y. Z., D. P., Y. J. and Q. X. conceived the project and designed the study. Y. Z., Y. P., J. B., C. X., C. L., H. H., and Q. X. performed experiments and analyzed the experimental data with inputs from D. P. and Y. J.. D. P. and Y. J. performed the numerical simulations and theoretical analysis with inputs from Y. Z. and Q. X.. Y. Z., D. P., Y. J., and Q. X. wrote the manuscript with inputs from all authors.

%
%
%
%



%

\end{document}